\documentclass[preprint,aps]{revtex4-1}
\usepackage[light,all,outline,bottomafter]{}
\usepackage{amsmath}
\usepackage{amsfonts}
\usepackage{amssymb}
\usepackage{epsfig}
\usepackage{verbatim}
\usepackage[sort&compress]{natbib}
\usepackage{graphicx}
\usepackage{lineno}

\usepackage{epstopdf}

\begin{document}
\linenumbers
\bibliographystyle{plainnat}

\title{Avalanches and Continuous Flow in Aging Aqueous Foam}
\author{Michael M. Folkerts}
\affiliation{Department of Physics, University of California at San Diego, La Jolla, California 92093, USA}
\author{Sam W. Stanwyck}
\affiliation{Department of Physics, University of California at San Diego, La Jolla, California 92093, USA}
\author{Oleg G. Shpyrko}
\affiliation{Department of Physics, University of California at San Diego, La Jolla, California 92093, USA}
\date{December 5, 2011}

\begin{abstract}
We used coherent light scattering in a multi-speckle detection scheme to investigate the mesoscale dynamics in aqueous foam. Time-resolved correlation of the scattered speckle intensities reveals the details of foam dynamics during aging. We introduce Temporal Contrast Analysis, a novel statistical tool that can be effective in characterizing structural rearrangements. Using Temporal Contrast Analysis we were able to detect two distinct dynamical components present during foam aging: spontaneous and intermittent, avalanche-like events and continuous, flow-like rearrangements in the foam structure. We were able to measure these contributions separately from the intrinsic statistical noise contribution, and thereby independently analyze the decay of each dynamical component during foam aging process.
\end{abstract}

\maketitle

\section{Dynamics of Jammed Systems}
The dynamics of densely packed grains, particles, droplets and bubbles are of great interest in the field of soft condensed matter, with examples ranging from granular piles and colloidal dispersions to foams, emulsions and biological tissues. As the density of particles is increased, these systems are often found to exhibit kinetic arrest. This kinetic arrest is characteristic of a ``jammed'' state in which the motion of any given particle is restricted due to the large number, and close proximity, of neighboring particles. As the system approaches a jammed transition, particle motion becomes more collective in nature, since in order to displace a single particle many other particles must move as well. As a result, many jammed systems exhibit slow, often non-equilibrium, relaxation commonly characterized by dynamical heterogeneities. This relaxation process is punctuated by intermittent, avalanche-like structural rearrangements \cite{cipelletti_slow_2005}, in contrast to continuous, flow-like (independent) motion of particles in the low density regime. \cite{bandyopadhyay_speckle-visibility_2005} \cite{durian_bubble-scale_1997} \cite{durian_foam_1995}
Here we investigate the non-equilibrium coarsening dynamics of aqueous foam using light scattering. Aqueous foam is one of the simplest systems exhibiting many characteristics of a jammed state. Despite being a combination of liquid and gas, foam retains many structural properties commonly found only in solids (for example, the ability to maintain its shape), as well as many properties of granular materials (such as flow under shear and jamming at low shear).
Aqueous foam is also an example of a non-equilibrium system which, once formed, undergoes a continuous coarsening process. During this process the average bubble size grows with aging time and the rearrangement rate of the tightly packed bubble structure decays over time. \cite{durian_bubble-scale_1997} \cite{durian_scaling_1991} \cite{pugh96} \cite{weaire93} \cite{hh97} \cite{hbraud_mode-coupling_1998} \cite{abate_topological_2007} \cite{gopal_relaxing_2003}

One of the main difficulties in directly coupling to intermittent, avalanche-like structural rearrangements \cite{abate_avalanche_2007} is that they are both temporally and spatially rare and unpredictable. We studied the dynamics of aqueous foam using a variation of Diffusing Wave Spectroscopy (DWS) where one records the changes in the coherent diffraction pattern (or speckle) formed by laser light scattered off the internal foam structure \cite{pine_diffusing_1988} \cite{pine_diffusing-wave_1990}. The coherent speckle pattern effectively becomes a ``fingerprint" of the specific configuration of scatterers (in this case the bubbles and the soap ridges forming the foam): a change in the configuration of scatterers  will modify the phases and intensities of interfering waves and therefore result in a different speckle pattern \cite{feng_correlations_1988} \cite{vera_angular_1996} \cite{dixon_speckle_2003}. Therefore DWS is a highly suitable technique for studying foam coarsening since it provides a direct way of coupling to the dynamics of a large ensemble of scatterers. \cite{cohen-addad_bubble_2001} \cite{cohen-addad_origin_2004} \cite{durian_multiple_1991}
\cite{earnshaw_diffusing-wave_1994} \cite{durian_scaling_1991}

\section{Materials and Methods}

The samples used here were Gillette Foamy Regular shaving foam. The typical dimension of the randomly distributed foam bubbles in this sample is on the order of a few micrometers, and the relatively long coarsening lifetime allows us to probe a substantial range of the dynamical life of the foam - from seconds to many hours. Due to the rapidly evolving nature of aging foam dynamics, we record many speckle intensity values simultaneously, a so-called ``multi-speckle" detection scheme \cite{kirsch_multispeckle_1996} \cite{wong_dynamic_1993}.

The coherent light source was a 2.0 mW HeNe laser, with a wavelength of 633 nm. The data were collected using a line scan charge-coupled device (CCD) detector. The 1024 pixel (pixel size 10 $\mu$m by 10 $\mu$m) spL2048-140k Basler line-scan CCD camera was placed 6.2 cm from the sample at an azimuthal scattering angle of 45 degrees. We collected 32 batches of data at 1 minute intervals, starting 1 minute and 30 seconds after we first expressed the foam. Each batch of data consisted of 32,768 line-scan frames with 74 $\mu$s exposure time taken 1 ms apart, for a total duration of 32.768 seconds for each batch.

\begin{figure}
\includegraphics[angle=0,width=1.0\columnwidth]{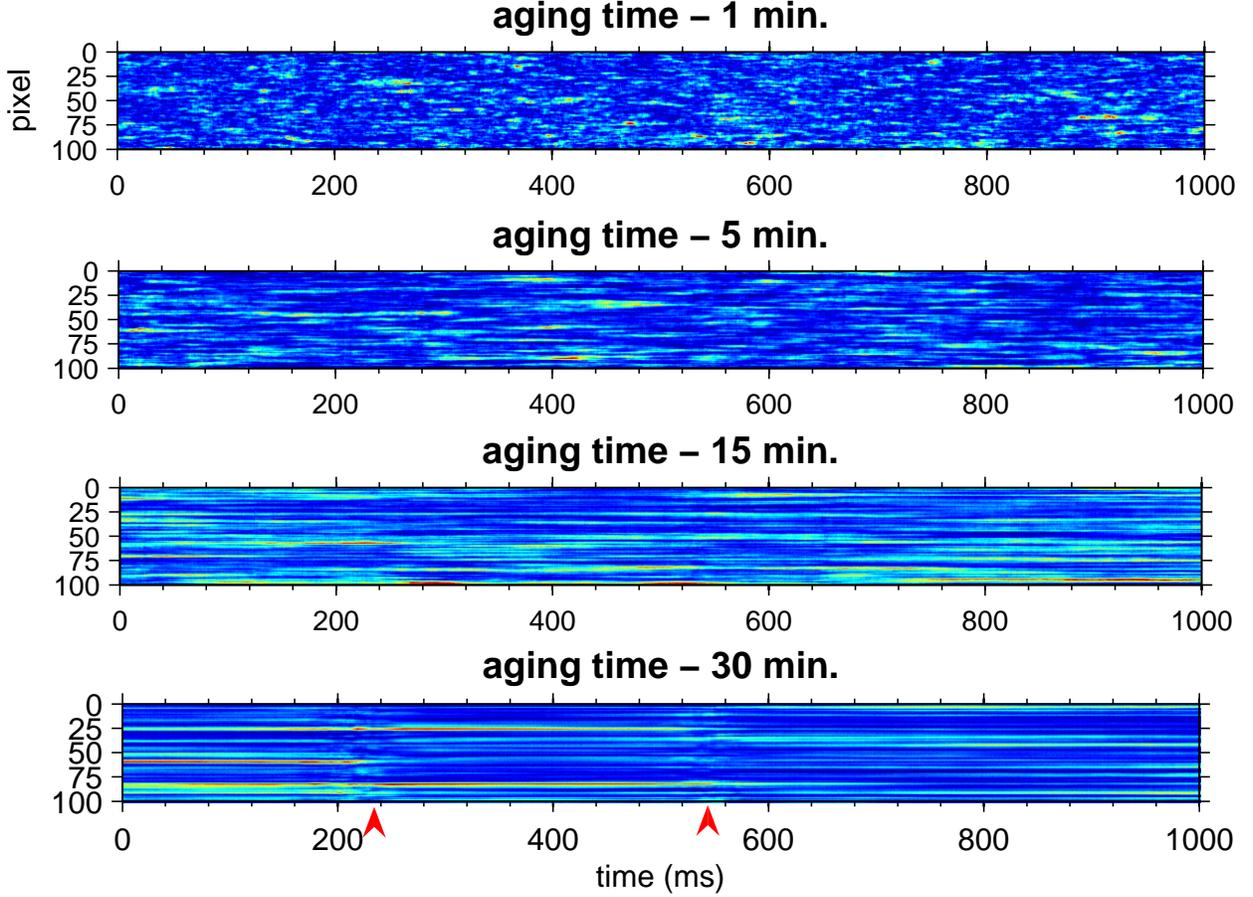}
\caption{Speckle intensity (color scale) for 100 pixel CCD region (vertical axis) as a function of time (horizontal axis, increasing from left to right) over a one second time interval. Red arrows indicate sudden changes in speckle patterns. The characteristic (exponential) decay time constant increases with aging time. Each image has been scaled for visibility.}
\label{fig:raw_data}
\end{figure}

\begin{figure}
\includegraphics[angle=0,width=1.0\columnwidth]{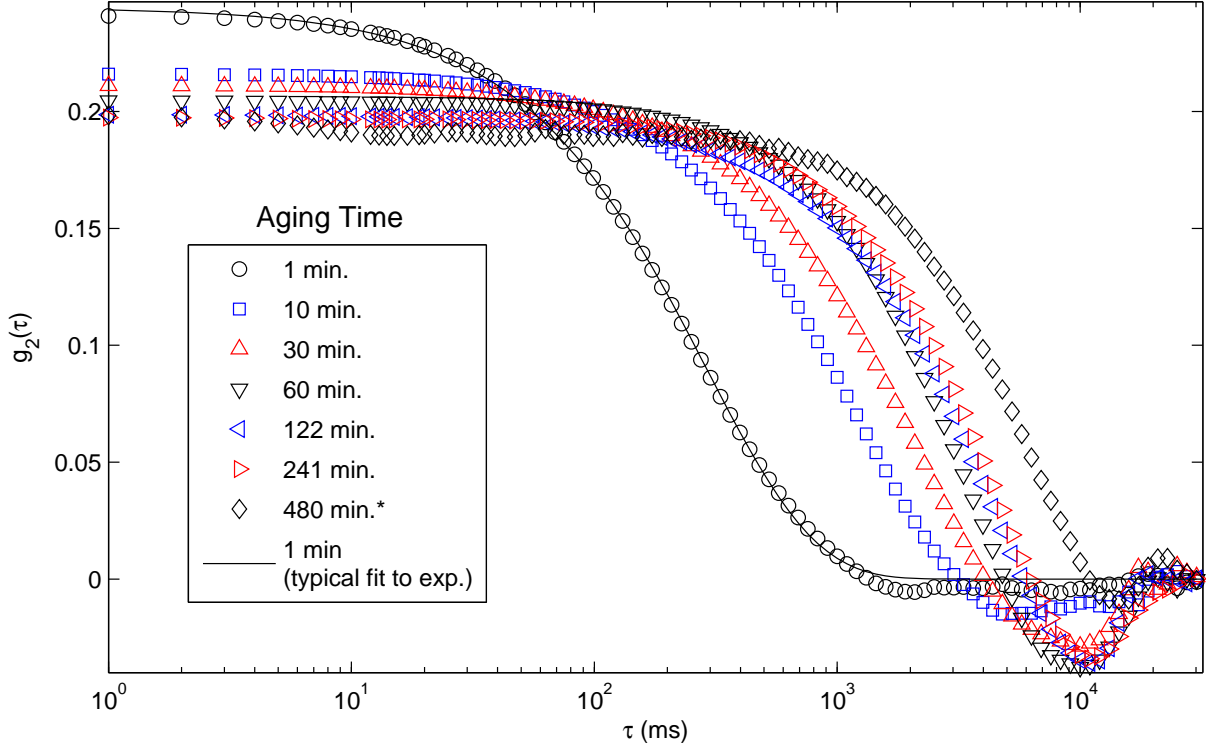}
\caption { Autocorrelation function at age 480 min. has been scaled by a factor of 5. }
\label{fig:autoCorr}
\end{figure}

\begin{figure}
\includegraphics[angle=0,width=1.0\columnwidth]{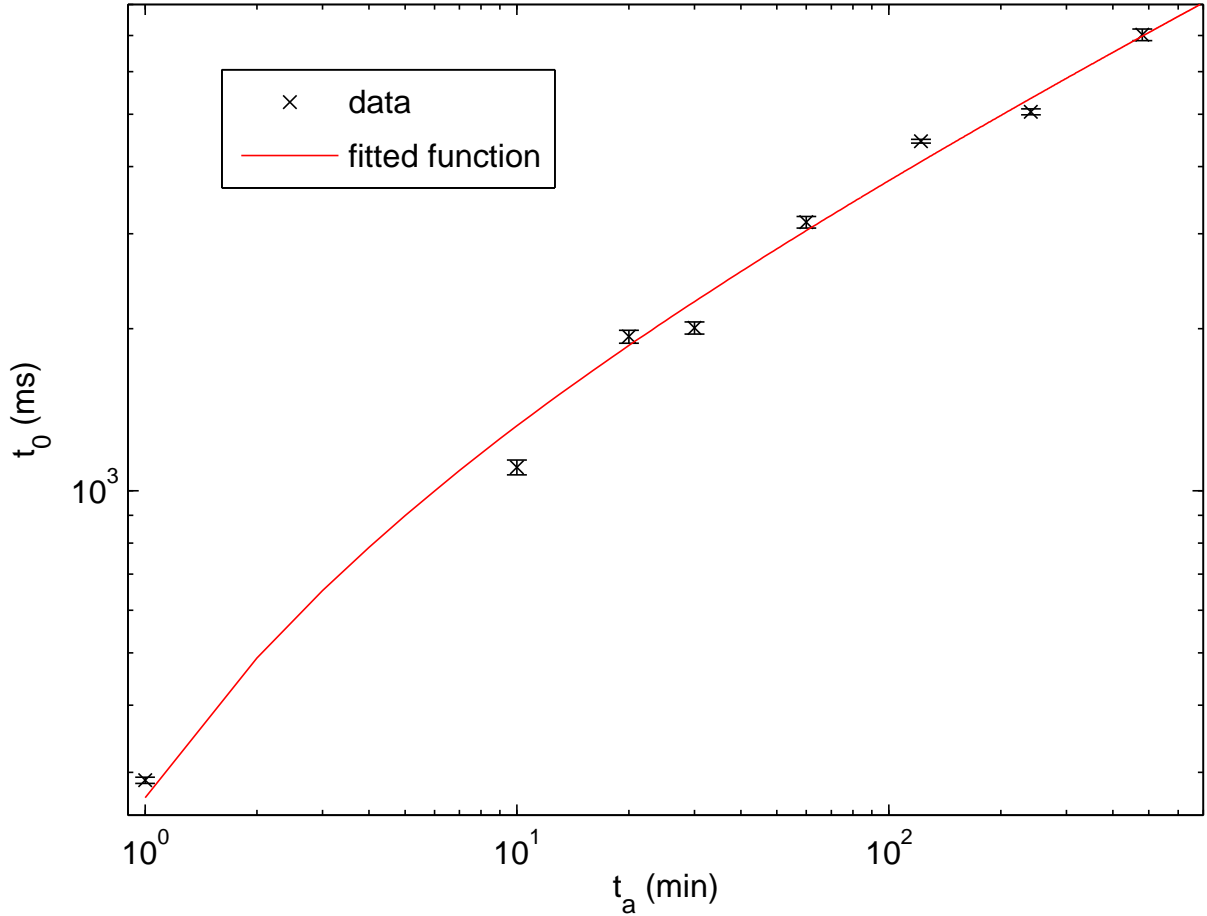}
\caption {Relaxation times $t_0$ obtained from fits of autocorrelation functions shown in Fig.~\ref{fig:autoCorr}, as a function of foam age. The best fit to the data is  $t_0 = a\cdot t_a^\gamma + b$ power law behavior, with $\gamma = 0.34$, or roughly a cube-root power law.} 
\label{fig:tauVSage}
\end{figure}

The raw speckle data is shown in Fig.~\ref{fig:raw_data}:intensity (indicated by color) over a 100 pixel CCD region  is plotted as a function of time, over a total time period of 1 second, with four different panels corresponding to selected foam aging  times of 1, 5, 15 and 30 minutes. For early aging times the speckle streak image appears chaotic due to the high number of rapid intensity fluctuations. As the foam ages the images show extended sections of constant speckle intensities, appearing as horizontal streaks in the image. These streaks represent the periods of time when the foam structure is fairly static. Periods of relative stability are occasionally interrupted by rapid rearrangements of speckle patterns; two examples of such intermittent rearrangements are identified with red arrows in the 30 minute aging time sequence (at approximately 240 and 550 ms).
We first quantified the aqueous foam dynamics by calculating the intensity autocorrelation function as a function of foam age. Intensity autocorrelation function is defined as:
\begin{eqnarray}
g_2(\tau) = \frac{\langle I(t)\cdot I(t+\tau)\rangle_t}{\langle I(t)\rangle^2}
\label{eq:autoCorr}
\end{eqnarray}
where I(t) and I(t+$\tau$) are speckle intensities recorded at time t and t+$\tau$, respectively, and averaging is performed over all accessible t and pixels while keeping $\tau$ constant.
Fig.~\ref{fig:autoCorr} shows the autocorrelation functions for 7 selected aging times (each was computed for a 32-second continuous batch dataset). As the sample ages, the characteristic decay time of the autocorrelation function increases, as  expected. Fig.~\ref{fig:tauVSage} shows the exponential relaxation time constants $t_0$ obtained from the fits of the autocorrelation data shown  in Fig.~\ref{fig:autoCorr} to a fitting model : $g_2(\tau)=a\cdot exp((\tau/t_0)^\beta)$. *, plotted as a function of aging time. Unlike the scaling behavior for very early ages (less than 1 minute), when the rate of bubble rearrangements during coarsening is found to scale as $t_a^{-2}$ \cite{durian_scaling_1991}, we observe a $t_0 \sim t_a^{-0.34}$ scaling over the range of aging times from 1 to 500 minutes.

\section{Two Time Correlation}
A standard approach to characterizing non-equilibrium dynamics involves calculating the two-time correlation function \cite{Fluerasu_twotime} (Eq.~\ref{eq:twoTimeCorr}), most conveniently displayed as a contour plot of $g_2(t_1,t_2)$ as a function of $t_1$ and $t_2$. As the dynamics of a system slow down (and characteristic relaxation time increases), the area of high correlation around the diagonal of the contour plot (where $t_1=t_2$) will broaden with time.

\begin{eqnarray}
g_2(t_1,t_2) = \frac{\langle I(t_1)\cdot I(t_2)\rangle_{q}}{\langle I(t_1)\rangle^2_{q}}
\label{eq:twoTimeCorr}
\end{eqnarray}

\begin{figure}
\includegraphics[angle=0,width=1.0\columnwidth]{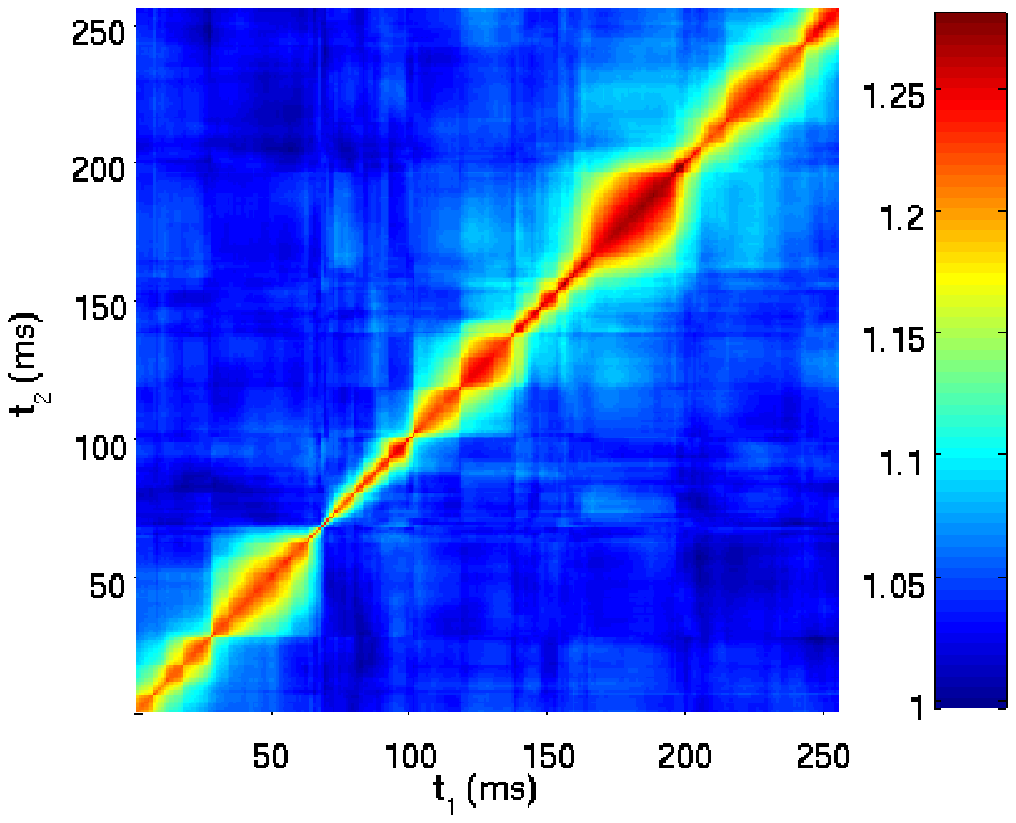}
\caption {Two-time correlation function $g_2(t_1,t_2)$  plotted for 250~ms time interval for 32 minute age.}
\label{fig:twoTimeCorr}
\end{figure}
Based on the similarities between Eq.~\ref{eq:autoCorr} and Eq.~\ref{eq:twoTimeCorr}, one would normally expect the monotonic broadening of $g_2(t_1,t_2)$ contours  in an aqueous foam system as the foam ages. However, the two-time correlation function shown in Fig.~\ref{fig:twoTimeCorr} exhibits a different behavior, with several regions of broadening with respect to $t_1=t_2$ diagonal, separated by  ``pinch-off" points where the contour abruptly narrows. This behavior is caused by a substantial intermittent dynamical component in foam samples: whenever there is a spontaneous (or intermittent) global rearrangement in the foam structure, the entire speckle pattern changes and the correlation to earlier speckle patterns becomes, suddenly, very poor. Following the rapid rearrangement, the $g_2(t_1,t_2)$ contours broaden again due to continuous dynamics,  until the next large intermittent rearrangement, at which point the contour of $g_2(t_1,t_2)$ pinches off again.
Although two-time correlation analysis fails to illustrate the diverging characteristic timescale in this particular case, the shape of $g_2(t_1,t_2)$ contours provides us with clear evidence of two distinct dynamic components of coarsening in aqueous foam, one continuous and another intermittent.

\section{Time Resolved Correlation}
Another method for coupling to the intermittent component of dynamics in a jammed system is Time Resolved Correlation (TRC) analysis \cite{cipelletti_slow_2005}. In contrast to Eq.~\ref{eq:autoCorr}, which takes a time average of correlation values at a fixed time offset $\tau$, TRC (Eq.~\ref{eq:TRC}) shows the correlation values for a single selected value of $\tau$ as a function of time, $t$.
TRC also corresponds to cross-sectional cut of the two-time autocorrelation function $g_2(t_1,t_2)$ along a 45-degree diagonal, offset from the $t_1=t_2$ line by $\pm\tau$.
\begin{eqnarray}
TRC(t) = \frac{\langle I(t)\cdot I(t+\tau)\rangle_{q}}{\langle I(t)\rangle_{q}\cdot\langle I(t+\tau)\rangle_{q}}
\label{eq:TRC}
\end{eqnarray}

\begin{figure}
\includegraphics[angle=0,width=1.0\columnwidth]{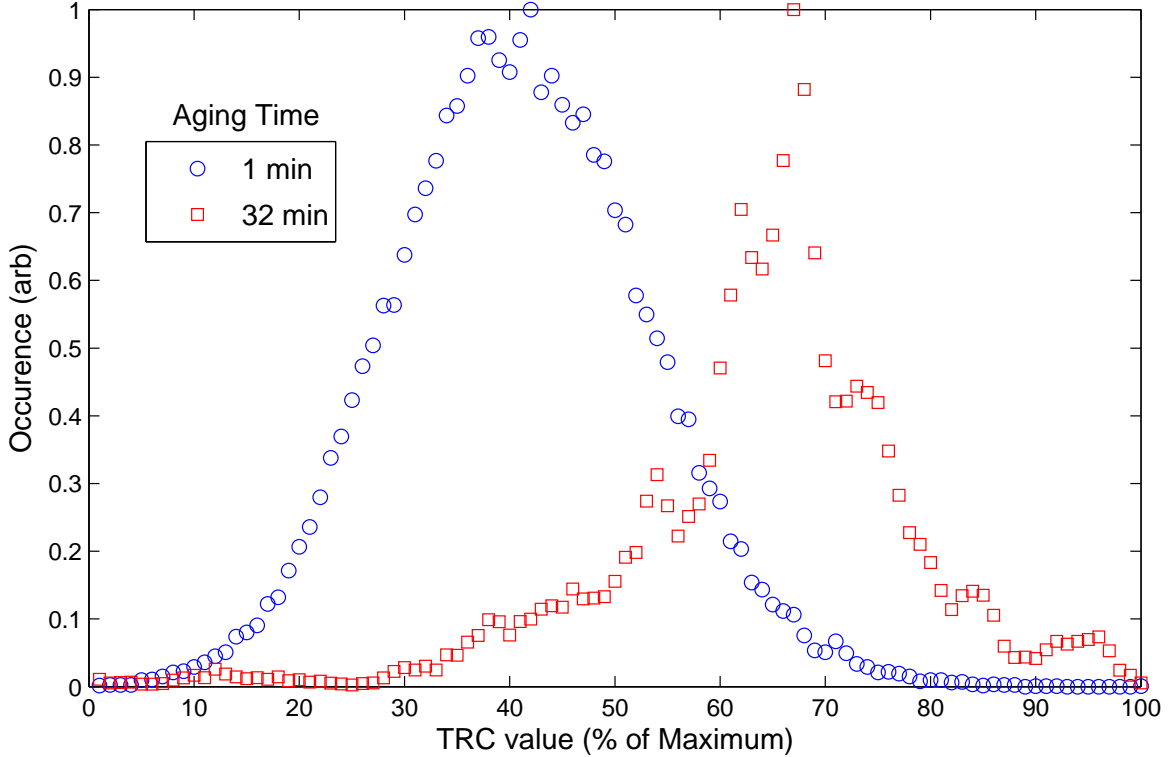}
\caption {The probability distribution function of TRC values becomes asymmetric, skewed towards lower values as the foam ages, due to temporal inhomogeneities.}
\label{fig:trcSkewness}
\end{figure}

For a dynamically homogeneous system the probability distribution of the TRC values is expected to follow a Gaussian distribution \cite{cipelletti_slow_2005}. The TRC of the foam data at early aging times (less than 5 minutes) displays a Gaussian distribution. This is because foam rearrangements are happening so rapidly that the dynamics closely resemble that of a continuous system. As the foam ages, an asymmetric tail emerges at the lower end of the TRC distribution [Fig.~\ref{fig:trcSkewness}]. The distribution becomes skewed towards lower TRC values as the dynamics become more temporally inhomogeneous.

\begin{figure}
\includegraphics[angle=0,width=1.0\columnwidth]{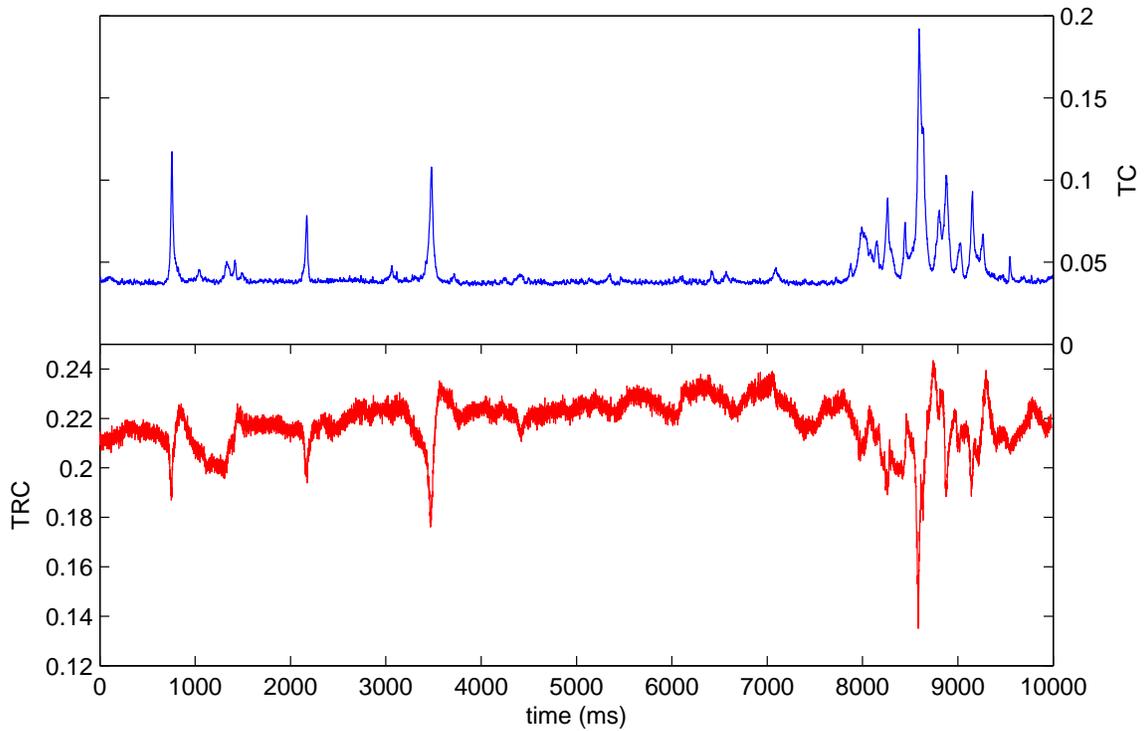}
\caption {Temporal Contrast (upper panel) and Time Resloved Correlation (lower panel) plotted over for 10 second period at a sample age of 31 minutes. Sudden structural rearrangements exhibit themselves as spikes in Temporal Contrast and coinciding dips in TRC data.}
\label{fig:trcANDtc}
\end{figure}

\section{Temporal Contrast}
While TRC provides clear evidence of heterogeneous dynamics with a substantial intermittent component, the gradually varying baseline of TRC makes it difficult to clearly identify the times at which intermittent rearrangements occur. In order to better isolate the intermittent component we devised a new method called Temporal Contrast (TC). The contrast (visibility) of spatially varying fringes is defined as the ratio of the variance of intensity with respect to the mean. We define a similar function as a ratio of variance to the mean for intensities measured within a specific pixel over a narrow time interval, $\tau$. We calculate the temporal contrast for each pixel, then average over all \textit{q} (pixels).
\begin{eqnarray}
T.Contrast(t,\tau)=\left\langle\frac{I_{max} - I_{min}}{I_{max} + I_{min}}\right\rangle_{q}
\label{eq:tContrast}
\end{eqnarray}
\begin{eqnarray}
I_{max/min}=max/min[I(t-\tau/2),
\cdots,I(t),\cdots,
I(t+\tau/2)] \nonumber
\label{eq:DefMinMax}
\end{eqnarray}
Where $\tau$ describes the duration of time interval over which the contrast is calculated.
In this study we used a $\tau$ = 20~ms window which was selected as a compromise between statistics and the temporal resolution. By taking the average over \textit{q}, continuous fluctuations are primarily suppressed and collective intermittent fluctuations show up as spontaneous spikes in Temporal Contrast.

\begin{figure}
\includegraphics[angle=0,width=1.0\columnwidth]{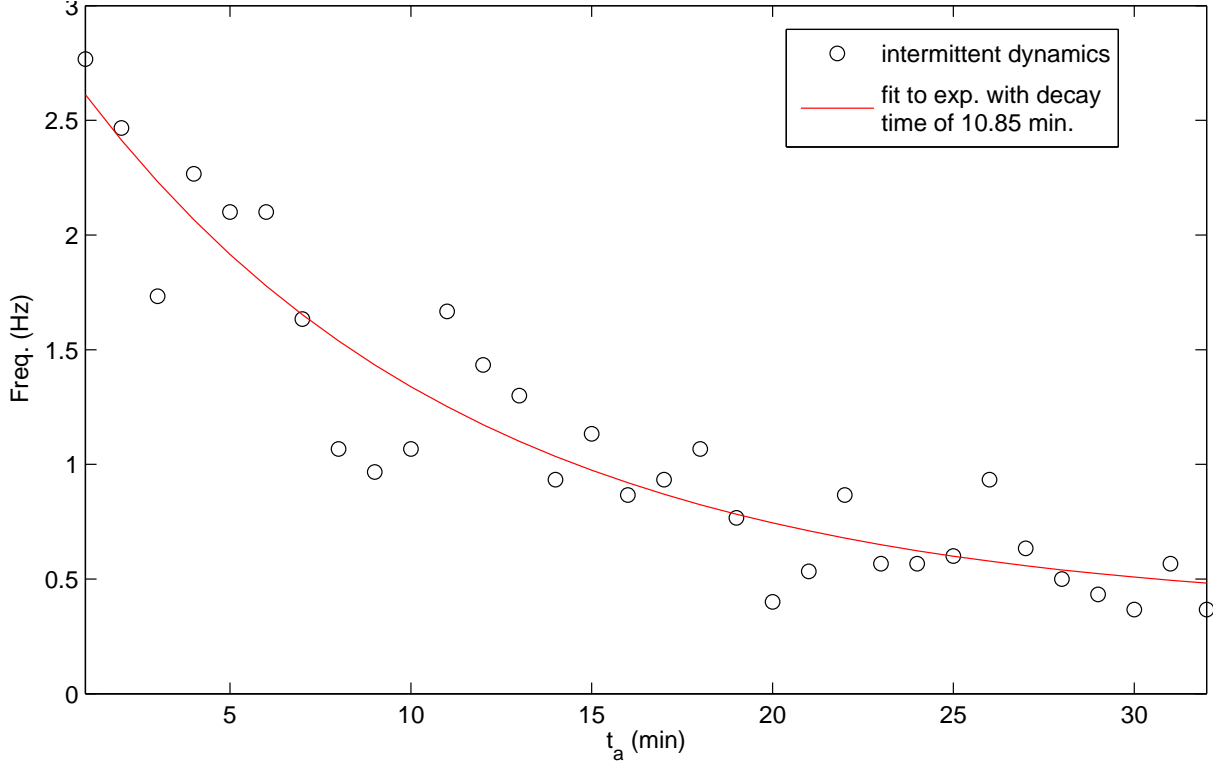}
\caption {The frequency of intermittent rearrangements as a function of aging time. The fit to the data shown with a line is represented by an exponential decay with a time constant $t_i = 10.85$ minutes. }
\label{fig:popsPerMin}
\end{figure}

The example of Temporal Contrast shown in Fig.~\ref{fig:trcANDtc}  provides a much more detailed picture of the sample dynamics - the spikes of various magnitudes correspond  to intermittent structural rearrangements that vary in either extent of the spatial dimensions or the relative displacements of the scatterers involved in the specific ``avalanche event.
As the foam coarsens the frequency of avalanche events clearly decreases and the Temporal Contrast value between the spikes stabilizes at a decreasingly lower, but finite baseline value.
Fig.~\ref{fig:popsPerMin}) shows the frequency of intermittent, ``avalanche-like rearrangements as a function of aging time. In calculating the avalanche frequency, we count the frequency at which Temporal Contrast value exceeded a certain threshold value, which was defined as the average between the maximum contrast (mean of 50 highest spike values) and the overall mean of the Temporal Contrast in each 32 second exposure.

One should note that if the sample structure is static for extended periods of time, the scattering speckle pattern is also static, resulting in Temporal Contrast value that should approach zero. This is not the case for Temporal Contrast calculated for our data:  for example, in Fig.~\ref{fig:trcANDtc}, the baseline value is about 0.04. There are two possible explanations for this finite value of Temporal Contrast: the first possibility is that finite photon counting statistics would lead to recorded intensity variations within each pixel even for absolutely static sample, resulting in non-zero value of Temporal Contrast, while the second possibility is that the sample undergoes slowly varying, continuous structural changes which results in (relatively small) speckle intensity fluctuations. As we demonstrate below, both of these effects are present in our sample.
When considering the effects of statistical noise from continuous dynamics on the Temporal Contrast baseline value, one should note that statistical variations of intensities from their mean are generally completely random as a function of time, while intensity fluctuations due to continuous dynamics will follow a continuous and monotonic trend (decreasing or increasing at various rates for different pixels), and this rate of change , measured over a period of time much shorter than the characteristic relaxation rate, will remain unchanged.
In order to separate these two types of contributions we down-sample the raw intensity datasets by a factor $n$, producing a sequence of frames consisting only of every $n$-th exposure, and discarding the exposures in between.
We then compute the Temporal Contrast baseline (defined as the mean of the 50 smallest Temporal Contrast values) in each of these down-sampled batches, and compare them as a function of down-sampling ratio $n$.

If in between spontaneous ``avalanche events the sample is truly static and the finite baseline value was due solely to the random noise present in the speckle, the down-sampling procedure will not change the baseline values of Temporal Contrast.
However, if the samples undergoes continuous structural evolution, resulting in continuous change in speckle intensities, the down-sampling will effectively compress these trends by a factor $n$, resulting in higher calculated baseline value of Temporal Contrast for these down-sampled datasets (note that Temporal Contrast will increase regardless of whether the intensity in the particular pixel is increasing or decreasing). For small values of Temporal Contrast (the relative changes in pixel intensities are small) the rate of intensity changes due to sample dynamics can be assumed constant, which would result in the baseline value of Temporal Contrast for down-sampled datasets increasing linearly with $n$.

\begin{figure}
\includegraphics[angle=0,width=1.0\columnwidth]{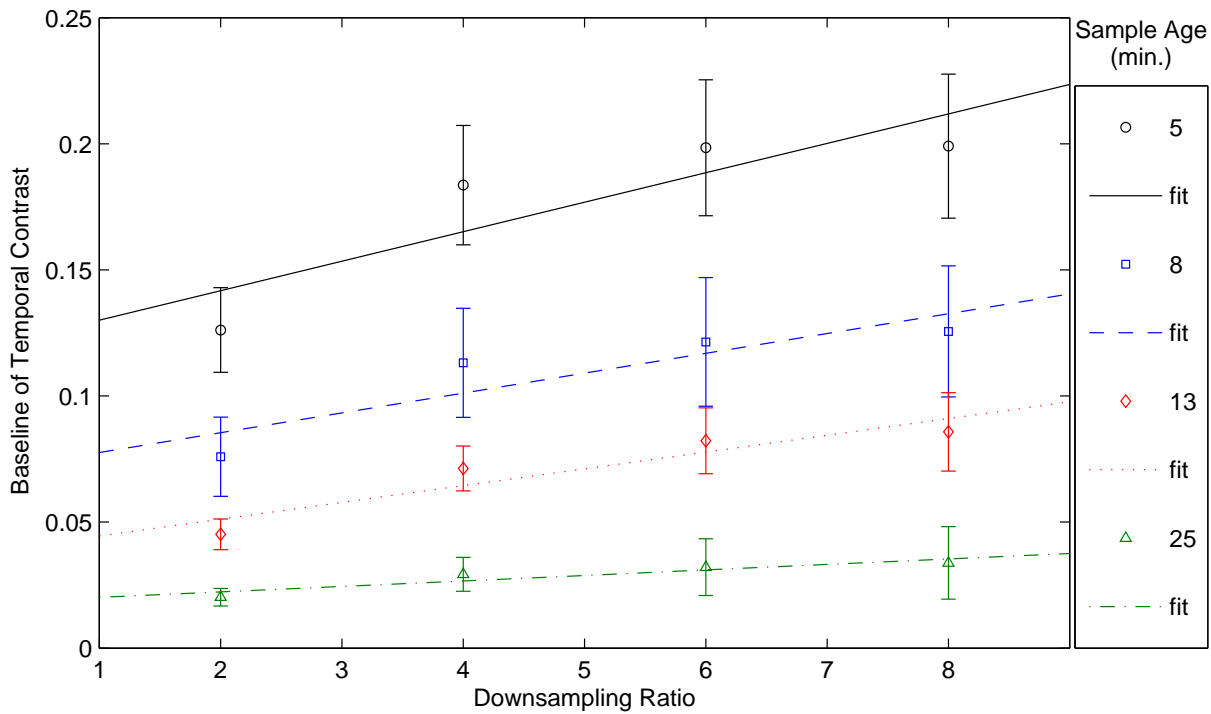}
\caption {The baseline value of Temporal Contrast calculated over each 32 second frame dataset for foam samples aged for 5, 8, 13 and 25 minutes as a function of down-sampling  ratio, $n$, shown along with linear fits. The slope of the fit represents the contribution of continuous dynamics component, while the constant term (y-axis intercept at n=0) is due to the random shot-noise statistical contribution.
}
\label{fig:extractCont}
\end{figure}
Fig.~\ref{fig:extractCont} shows the linear dependence of the Temporal Contrast baseline as a function of down-sampling ratio, $n$, for different aging times. By measuring the slope of the data with respect to $n$, we can directly measure the change in the continuous dynamics as the sample is aging â as can be seen from Fig.~\ref{fig:extractCont}, the slope decreases during aging, which implies that the rate at which the sample undergoes continuous evolution is also decreasing over time.
\begin{figure}
\includegraphics[angle=0,width=1.0\columnwidth]{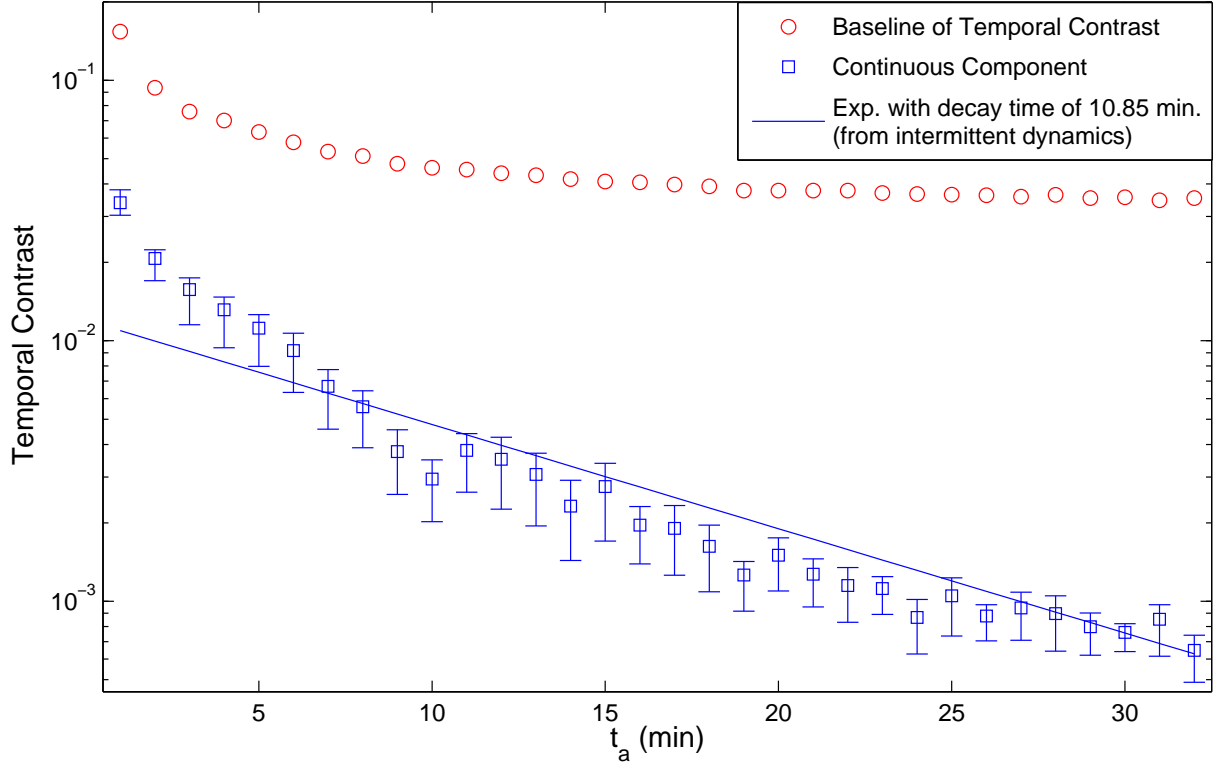}
\caption {The change in baseline value of Temporal Contrast (red circles) and the rate of continuous dynamics (blue circles) obtained from fitting the slope of change in baseline for down-sampled data, as shown in Fig.~\ref{fig:extractCont} , as a function of aging time, $t_a$.  The difference between the two curves is the contribution to the baseline value of Temporal Contrast due to shot-noise fluctuations..}
\label{fig:contDyn}
\end{figure}
Figure \ref{fig:contDyn} shows the baseline Temporal Contrast, as well as the rate of continuous dynamics level as a function of sample age. The difference between the baseline and the continuous component is the contribution to the baseline due to finite counting statistics, or shot-noise. The continuous component of the dynamics decays exponentially over time, with a time constant comparable to the exponential decay time constant of the rate of avalanche-like intermittent rearrangements, shown in Fig.~\ref{fig:popsPerMin}.  This remarkable similarity in behavior of intermittent and continuous dynamics may be indicative of a coupling between the spontaneous and rare avalanche-like (temporally heterogeneous) events and the slow continuous flow that occurs in between avalanches which corresponds to the temporally homogeneous component of the dynamics. One possible model of the coupling mechanism is that the flow-like motion occurs in response to stress fields and voids created as a result of sudden and drastic rearrangements of the foam structure that occur during and preceding an avalanche event. However, it is also possible that the flow-like continuous rearrangements can themselves lead to building up of stresses that are ultimately culminated in a rapid avalanche event or series of avalanches, due to, for example, the bursting of soapy ridge or multiple ridges.

\section{Summary/Conclusions}
By using several  traditional statistical methods of speckle intensity fluctuation analysis, we were able to show that dynamics in foam exhibit slowing behavior as the foam ages and coarsens. We also introduced a new method of studying these dynamics in foam, Temporal Contrast, which is especially effective in elucidating the intermittent, spontaneous structural rearrangements in systems consisting of a large number of scatterers. Through the statistical down-sampling analysis of Temporal Contrast baseline values we were able to advance upon previous studies of foam dynamics by showing that in between intermittent rearrangement the foam sample undergoes continuous flow-like dynamics. The rate of this slow, continuous, dynamical component can be separated from the contribution of statistical noise to the Temporal Contrast data, and was shown to decay exponentially during sample aging, in striking similarity to exponential decay of the rate of intermittent rearrangements as a function of aging time. This may indicate that the continuous, flow-like rearrangements are coupled to the intermittent, avalanche-like events, even though the precise nature of coupling remains unknown. One can speculate, however, that the continuous flow-like dynamics may be due to the slowly varying internal stresses in response to foam drainage, while avalanche-like rearrangements are due to the bursts of the film connecting the bubbles.

The statistical analysis methods of temporal speckle fluctuations applied by us to study foam coarsening will be useful for  studying complex non-equilibrium dynamics in a wide variety of other systems, where both intermittent and continuous dynamics may be present. The examples of such systems include not only a wide range of soft matter and granular systems near jamming transitions, but also a wide range of hard condensed matter systems, such as dynamics of magnetic, ferroelectric, charge-ordered domains as well as systems in the vicinity of electronic, structural and magnetic phase transitions. The analysis of temporal fluctuations in visible light speckle is directly applicable to shorter length scales probed by techniques such as X-ray Photon Correlation Spectroscopy and Fluorescence Intensity Fluctuation Spectroscopy.

\section{Acknowledgements}
This work at UC San Diego was supported by NSF CAREER Grant No. 0956131. O.~S. acknowledges support from Hellman foundation.
\bibliographystyle{unsrt}
\bibliography{foam-1}
\end{document}